\documentclass[10pt]{wlscirep}
\usepackage[utf8]{inputenc}
\usepackage[T1]{fontenc}
\usepackage{lineno}
\usepackage{amsmath,amssymb}
\DeclareMathAlphabet{\mathcal}{OMS}{cmsy}{m}{n}
\title{Emergence of topological superconductivity in the presence of chiral magnetism in 2D CrInTe$_3$}

\author[1,$\dag$]{Arnob Mukherjee}
\author[2]{Fengyi Zhou}
\author[1]{Soheil Ershadrad}
\author[3]{Tanay Nag}
\author[2]{Duo Wang}
\author[1]{Biplab Sanyal}
\affil[1]{Department of Physics and Astronomy, Uppsala University, Box-516, 75120 Uppsala, Sweden}
\affil[2]{Faculty of Applied Sciences, Macao Polytechnic University, Macao SAR, 999078, China}
\affil[3]{Department of Physics, BITS Pilani-Hyderabad Campus, Telangana 500078, India}
\affil[$\dag$]{corresponding author(s): Arnob Mukherjee (arnob.mukherjee@physics.uu.se)}

\begin{abstract}
We propose a general framework for designing a two-dimensional (2D) topological superconductor (TSC) using a magnet-superconductor hybrid system. This setup involves a monolayer of CrInTe$_3$, which hosts noncoplanar magnetic textures, in proximity to a 2D $s$-wave superconducting layer. Serving as an alternative to $p$-wave superconductors, this configuration induces a topological superconducting phase and is a promising platform for realizing the 2D Kitaev model, which supports Majorana zero-energy modes through emergent $p$-wave symmetry superconducting pairing. Notably, the magnetic moments break time-reversal symmetry while the superconducting state preserves particle-hole symmetry, placing our system in the Altland-Zirnbauer class $D$ and ensuring robust Majorana excitations. We first perform density functional theory-based simulations to study a monolayer of CrInTe$_3$, from which essential magnetic characteristic parameters, such as Heisenberg exchange interaction and Dzyaloshinskii-Moriya interaction (DMI), are calculated using the state-of-the-art Liechtenstein-Katsnelson-Antropov-Gubanov (LKAG) approach. With a substantial DMI coupling exhibited in CrInTe$_3$, large-scale Monte Carlo simulations reveal the stabilization of a noncoplanar spiral magnetic state as ground state. In this magnetic phase, we observe a transition from corner modes in the zero-energy local density of states (LDOS) to edge modes as the chemical potential ($\mu$) varies. Furthermore, under a finite magnetic field, the system enters a mixed magnetic state, characterized by isolated skyrmions and spiral domain walls, which lead to unique low-energy localization of electronic wave functions, rendering the system an insulator. Finally, we discuss potential experimental realizations of TSC in this magnet-superconductor interfacial system, using real-space probes such as scanning tunneling microscopy (STM).
\end{abstract}
\begin{document}

\flushbottom
\maketitle

\thispagestyle{empty}

\section*{Introduction}

Since its discovery, superconductivity has been an inspiration for many novel concepts and applications \cite{bednorz1986, kamihara2008}. Topological superconductors (TSCs), which host charge-neutral Majorana fermion quasiparticles, have recently garnered significant attention, primarily due to their potential applications in quantum computing. \cite{lutchyn2018, sato2017, Elliott2015, ando2015}. In TSCs, Majorana zero modes (MZMs) that obey non-Abelian exchange statistics are of particular interest: such exotic kinds of particles can be trapped either in zero-dimensional defects or can unfold in one-dimensional non-singular defects, leading to new ways of quantum manipulation \cite{Schnyder2008, Altland1997, Teo2010, ryu2010}. In TSCs, inversion-symmetry-breaking induced spin-orbit coupling (SOC) is important as it breaks spin degeneracies and, in turn, provides a platform for engineering a topological superconducting phase \cite{mourik2012, deng2012, nadj2014, das2012, Ruby2015, pawlak2016}. It has been experimentally observed in quasi-one dimensional semiconductor-superconductor hybrid nanowire systems that the external Zeeman field lifts the residual Kramers' degeneracies, allowing the appearance of MZMs.

However, there has been a surge of interest in fabricating hybrid 2D systems with magnet-superconductor (M/SC) interfaces, which could be a fertile ground for various kinds of TSC. The magnetic textures such as spirals, chiral magnetism, and skyrmions simultaneously generate the desired SOC and necessary gauge fields that will make the induction of TSC possible at close proximity. For example, this interplay between magnetic textures and superconductivity has been experimentally demonstrated to underlie the importance of stability for topological superconductivity \cite{palacio2019, maggiora2024}. In the case of magnetic impurities on $s$-wave superconductors, the interaction has been extensively studied in the context of topological superconductors (TSCs) that host Majorana zero modes (MZMs). This interaction leads to the formation of Yu-Shiba-Rusinov bands \cite{Pientka2013, shiba1968} within the superconducting gap. These bands facilitate phase transitions analogous to those in the one-dimensional Kitaev model, resulting in the emergence of MZMs \cite{kitaev2001, kitaev2009}.

The realization of $p$-wave superconductivity with higher Chern numbers and dispersive chiral Majorana edge modes is especially remarkable in two-dimensional extensions of this model \cite{Zhu2018, Yan2018, wang2018prl, RXZhang2019, Volpez2019, 1ArnobG2024}. This crucial dimensional transition realizes non-dispersive Majorana flat edge modes localized at the edges of 2D domains detectable by scanning-tunneling microscopy and angle-resolved photoemission spectroscopy. Through the combined effect of spin-spiral order and $s$-wave superconductivity, a topologically nontrivial $p$-wave–like phase can emerge, evolving into a gapless 2D TSC upon gap closure. This phenomenon may manifest in transition metal monolayers deposited on $s$-wave superconductor substrates, as previously discussed \cite{nadj2014, schneider2019}.

Frustrated lattice geometries, such as triangular lattices, also offer exciting opportunities for exploring exotic phases. The interplay between strong electron correlations, Rashba SOC, and geometrical frustration can yield spin liquids, noncoplanar magnetic states, or partially ordered magnetic textures. For instance, CrInTe$_3$ crystallizes in a quasi-two-dimensional triangular lattice with a distinct five-layer Janus configuration and belongs to the $\text{C}_{\text{s}}$ point group symmetry \cite{zhou2024}. In this lattice, varying Cr–Te bond lengths and Cr–Te–Cr bond angles introduce geometrical frustration, enriched further by the presence of SOC and Dzyaloshinskii–Moriya interaction (DMI). As a result, magnetic spiral states or nontrivial chiral orders can develop, providing a natural setting for topological superconductivity when placed in proximity to an $s$-wave superconductor.

Crucially, topological superconductors emerging from magnet–superconductor hybrids often break time-reversal symmetry (TRS) while retaining particle-hole symmetry (PHS), placing these systems in Altland-Zirnbauer Class $D$ \cite{Chiu2013, Fu2023, Langbehn2017, Cornfeld2019}. Class $D$ describes superconductors in which magnetic exchange or other magnetic effects break TRS, yet the intrinsic structure of the Bogoliubov–de Gennes Hamiltonian preserves PHS—a condition underpinning the appearance of robust Majorana modes. As we detail in the following sections, these ideas come together in a comprehensive framework for realizing and probing TSC in frustrated magnet–superconductor interfaces, with the precise model presented in the Model section.

\begin{figure}[ht]
  \centering
  \includegraphics[width=0.8\textwidth,clip=true]{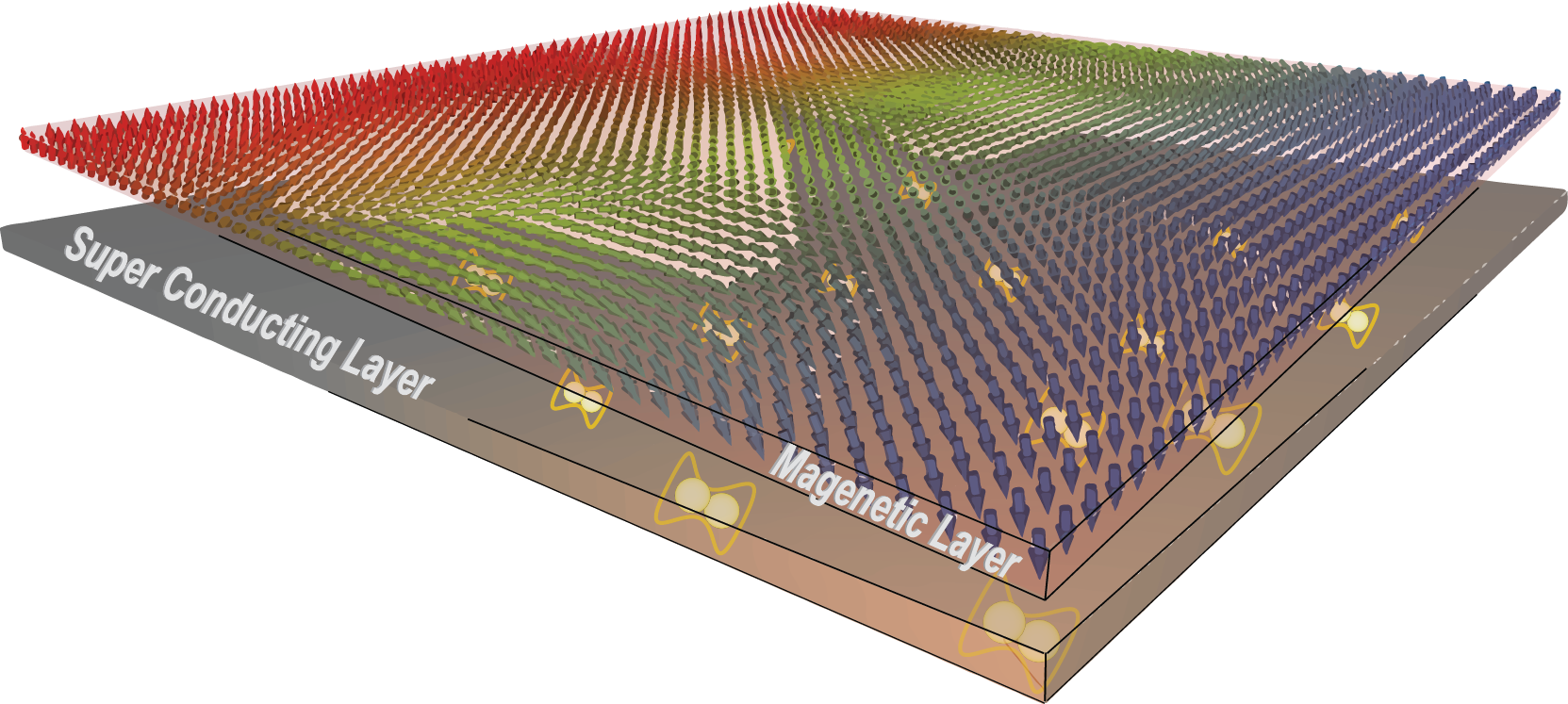}
  \caption{\textbf{Schematic setup of the model.} %A two-dimensional (2D) triangular lattice of CrInTe$_3$ with a spin spiral state is placed on the surface of an $s$-wave superconductor. The interplay between the magnetic texture and superconductivity enables the realization of robust topological phases.
  A two-dimensional (2D) magnetic moolayer of CrInTe$_3$ with a spin spiral state is placed on the surface of an $s$-wave superconductor. The interplay between the magnetic texture and superconductivity enables the realization of robust topological phases.}
  \label{fig:schematic}
\end{figure}

In this work, we explore the hybridization of nontrivial magnetic phases in CrInTe$_3$ with $s$-wave superconductivity on a triangular lattice to investigate the emergence of topological superconducting phases and Majorana modes. Figure \ref{fig:schematic} presents a schematic of our model, which consists of a heterostructure where a monolayer of CrInTe$_3$ with a spin spiral state is placed atop an $s$-wave superconducting layer in close proximity. This strategy dovetails with recent theoretical and experimental advances, emphasizing the intimate interplay among magnetic textures, superconductivity, and complex lattice geometry. We also highlight open questions relevant for designing robust TSC materials, especially in regard to real-space probes, such as scanning tunneling microscopy, that can reveal zero- or low-energy states localized at edges or defects. By elucidating these mechanisms within intricate lattice models, our results serve the broader quest to identify practical, stable platforms for topological quantum computation.

\section*{Results}

\subsection*{Model}

We consider a minimal electronic model Hamiltonian in real space for a 2D (M/SC)  heterostructure system, where a magnetic layer is deposited on the surface of a conventional $s$-wave superconductor:

\begin{center}
\begin{eqnarray} \label{eq:Ham_exchange_SC}
  H &=& -t \sum_{\langle ij \rangle, \alpha} (c^{\dagger}_{i\alpha} c^{}_{j\alpha} + H.c.) - \mu \sum_{i, \alpha} c^{\dagger}_{i\alpha} c^{}_{i\alpha} + J \sum_{i, \alpha \beta} c^{\dagger}_{i\alpha} ({\bf{S}}_i \cdot {\boldsymbol{\sigma}}_i)_{\alpha \beta} c^{}_{j\beta} + \Delta_s \sum_{i} (c^{\dagger}_{i\uparrow} c^{\dagger}_{i\downarrow} + H.c.)
\end{eqnarray}
\end{center}

\noindent
Here, $c^{\dagger}_{i\alpha}$ creates an electron at lattice site $i$ with spin $\alpha$, and ${\boldsymbol{\sigma}}$ is the vector of Pauli spin matrices. We denote the chemical potential as $\mu$, and $t$ represents the hopping amplitude between nearest-neighbor sites. We set the hopping parameter $t=1$ as the reference energy scale of the system. The superconducting pairing amplitude is set to $\Delta_s = 0.5$, consistent with conventional $s$-wave superconductors. The moments $\{ {\bf S}_i \}$ of the magnetic Cr atom spin at site $i$ in the magnetic layer coupled to the conduction electrons in the superconducting layer via a Hund's interaction of magnitude $J$, are described by the third term in the Hamiltonian. Since the full superconducting gap suppresses Kondo screening, we treat the local moments as classical unit vectors with unit length $|{\bf S}|=1$.

%This generic form of Hamiltonian underlines the purpose of magnetic textures to drive pseudo-spin-orbit coupling on the M/SC interface, an ingredient for stabilizing the TSC phase.

This generic form of the Hamiltonian underscores how magnetic textures at the M/SC interface can induce pseudo-spin-orbit coupling, a crucial factor for stabilizing the topological superconducting (TSC) phase. The $\{ {\bf S}_i \}$ represent all the spin textures in Eq. \ref{eq:Ham_exchange_SC}, arising as the magnetic ground state of Cr ions. The magnetic moments are defined as ${\bf S}_i = |{\bf S}| (\sin \theta_i \cos \phi_i, \sin \theta_i \sin \phi_i, \cos \theta_i)$, locally mimicking a variation of the magnetic impurities. These angles $\theta$ and $\phi$ can be extracted from spin textures generated by Monte Carlo simulations for M/SC systems. Note that, unlike earlier studies done in 2D MSH systems, this Hamiltonian does not contain intrinsic Rashba Spin-Orbit (RSO) interaction \cite{Klinovaja2013, herbrych2021, Fu2008, Hui2015, Mashkoori2019}. 

Although we have already introduced the classification of our system in Class $D$ in the Introduction, it is instructive to see how this arises directly from the Hamiltonian in Eq. \eqref{eq:Ham_exchange_SC}. First, the on-site $s$-wave pairing term $\Delta_s \sum_{i} (c^{\dagger}_{i\uparrow} c^{\dagger}_{i\downarrow} + H.c)$ enforces particle-hole symmetry (PHS), a fundamental property of Bogoliubov–de Gennes (BdG) Hamiltonians. Meanwhile, the coupling ${\bf{S}}_i \cdot {\boldsymbol{\sigma}}_i$ breaks time-reversal symmetry (TRS), since a static, nonzero magnetic texture $\{ {\bf S}_i \}$ reverses sign under time reversal. Consequently, although PHS is preserved, TRS is absent, and thus there is no chiral symmetry (which would require both PHS and TRS). As a result, this system belongs to Altland-Zirnbauer Class $D$. This symmetry classification underpins the possibility of realizing Majorana bound states in magnet–superconductor heterostructures.

The lattice Hamiltonian in Eq. \ref{eq:Ham_exchange_SC} can be expressed in the Nambu basis $\Psi_i = \{ {c_{i\uparrow}, c_{i\downarrow}, c^{\dagger}_{i\downarrow}, -c^{\dagger}_{i\uparrow}} \} ^\text{T}$ as:
\begin{eqnarray} \label{eq:Ham_exchange_SC_Numbu}
    H &=& \sum_{\langle ij \rangle} c^{\dagger}_{ij} \{ \mu\tau_z\sigma_0 + \Delta_s\tau_x\sigma_0 + J ({\bf S} \cdot  \boldsymbol{\sigma})\tau_0\} c^{}_{ij} - t \sum_{\langle ij \rangle} c^{\dagger}_{ij} \tau_z\sigma_0 (c^{}_{i+1j} + c^{}_{ij+1})  + H.c.
  \end{eqnarray}

  \noindent
The Pauli vector operators $\boldsymbol{\sigma}$ and $\boldsymbol{\tau}$ act on spin states $(\uparrow,\downarrow)$ and particle-hole configurations, respectively. $\tau_{1}\sigma_{2}$ denotes the Kronecker product $\tau_{1}\otimes\sigma_{2}$.

We have extracted the material-specific exchange parameters for monolayer CrInTe$_3$ using \textit{ab initio} electronic structure calculations based on the Density Functional Theory (DFT). These parameters subsequently optimized the ground-state spin texture by solving extended Heisenberg models through large-scale Monte Carlo (MC) simulations. The resulting ground-state spin configurations serve as the magnetic layer in our M/SC hybrid system. This setup is experimentally feasible, as interface engineering has already proven essential in characterizing the topological Hall effect in chiral magnetic materials. Recent experimental studies, such as the observation of a 3Q magnetic state on the surface of an Mn/Re (0001) $s$-wave superconductor, suggest the potential for realizing topological superconductivity in similar systems \cite{Wiesendanger2020}.

\subsection*{DFT results}
Monolayer CrInTe$_3$ is a Janus material consisting of five atomic layers and belongs to the $C_s$ point group. The structural model was built with a $20 ~\text{\AA}$ vacuum in the direction perpendicular to the surface. A $2\times2\times 1$ supercell was employed for simulating different magnetic states in the collinear limit. Geometry optimization and determination of the lowest energy magnetic state were performed using the plane wave basis set and projector-augmented wave code VASP~\cite{kresse1996efficiency}. The obtained crystal structure and magnetic configuration were then used as input in the full-potential linear muffin-tin orbital (FP-LMTO) code RSPt~\cite{wills2010full}, where more accurate self-consistent electronic calculations were performed.
By introducing small perturbations into the spin system (so-called force theorems~\cite{liechtenstein1987local,kubler2021theory}), non-self-consistent calculations were conducted, and essential characteristic magnetic parameters, such as the Heisenberg exchange parameters $\mathcal{J}_{ij}$, Dzyaloshinskii-Moriya interaction vectors $D_{ij}$, and single-ion anisotropy constant $K$, were obtained.
In this part, the generalized gradient approximation (GGA)~\cite{perdew1996generalized} was used to describe the exchange-correlation effect, and the on-site Hubbard U correction (U$_{\rm eff} = 3$ eV)  was augmented to have a better description for the strongly correlated Cr $d$-electrons~\cite{dudarev1998electron}. The plane-wave energy cutoff was set to 520 eV, and the k-mesh integration for the first Brillouin zone was chosen as $9\times 9 \times 1$ and $15\times 15\times 1$, with convergence criteria of 0.005 eV/{\AA} and $1\times 10^{-10}$ Ry for geometry optimization and self-consistent calculations, respectively.

In this collinear limit, CrInTe$_3$ exhibits ferromagnetism (FM) as the lowest energy magnetic state, with lattice parameters of 8.56 {\AA} for both $a$ and $b$. The calculated magnetic parameters are shown in Table~\ref{tab:m_parameters}.

\par
An effective spin Hamiltonian was then constructed as follows
\begin{equation}
    \centering
	\label{eq:Hamiltonian}
	% \begin{split}
		\mathcal{H}_{\rm eff}=  -\sum_{ i, j}\mathcal{J}_{ij}\, \mathbf{S}_i \cdot \mathbf{S}_j - \sum_{ i, j} \mathbf{D}_{i j} \cdot\left(\mathbf{S}_i \times \mathbf{S}_j\right)  - \sum_i\ K_i \left(\mathbf{S}_i^z\right)^2  -\sum_i \textbf{B}^{ext}\cdot\mathbf{S}_i\, .
	% \end{split}
\end{equation}
In the above expression, the four terms represent isotropic Heisenberg bilinear exchange, Dzyaloshinskii-Moriya interaction, single-ion anisotropy and Zeeman terms respectively.
Based on this Hamiltonian, a $50\times 50\times 1$ supercell with periodic boundary conditions was built, and a series of classical Monte Carlo simulations were performed to obtain the magnetic equilibrium states at different temperatures and external magnetic fields. To ensure the system reaching its equilibrium state, an annealing process with gradually decreased temperature and $100,000$ simulation steps was conducted.
The results show that, at 0 K and 0 T, the magnetic ground state of CrInTe$_3$ exhibits a spin spiral state, with a domain phase of 40 degrees and a propagation periodicity of 2.22 nm. With an increasing external magnetic field up to 5.5 T, the stripe-like magnetic phase breaks into shorter pieces, forming elliptical skyrmions. A stronger magnetic field further decreases the major axis of the spin swirling morphology, starting to form a circular spin state at 6.1 T.

\subsection*{Realization of TSC in Spiral state}
% Moving towards the numerical results associated with the spin spiral state of CrInTe$_3$ [see Fig. \ref{fig:spinspiral_ssf} (a)] at zero magnetic field $(B = 0$ T), we first analyze by computing the static structure factor (SSF),
% \begin{eqnarray}
%   \centering
%   S_f({\bf q}) &=& S^{x}_f({\bf q}) + S^{y}_f({\bf q}) + S^{z}_f({\bf q}), \nonumber \\
%   S^{\mu}_{f}({\bf q}) &=& \frac{1}{N^2} \bigg \langle \sum_{ij} S^{\mu}_i S^{\mu}_j~ e^{-{\rm i}{\bf q} \cdot ({\bf r}_i - {\bf r}_j)} \bigg \rangle
%  \label{eq: SSF}
%  \end{eqnarray}
% \noindent
% where, $\mu = x, y, z$ denotes the components of the spin vector and ${\bf r}_i$ is the position vector for spin ${\bf S}_i$. Spiral state with Bragg peak located at $\pm {\bf q}$ for one specific $\bf q$ in the 1st Brillouin zone (1BZ) [see Fig. \ref{fig:spinspiral_ssf} inset]. For the single-Q spiral state, one pair of $K$ points is spontaneously selected from three equivalent choices. These Bragg peaks in reciprocal space are probed using neutron scattering experiments. Now we study the effect of this spiral state on the superconducting layer by solving the Hamiltonian in Eq. [\ref{eq:Ham_exchange_SC_Numbu}].

\begin{figure}[ht]
  \centering
  \includegraphics[width=0.6\textwidth,clip=true]{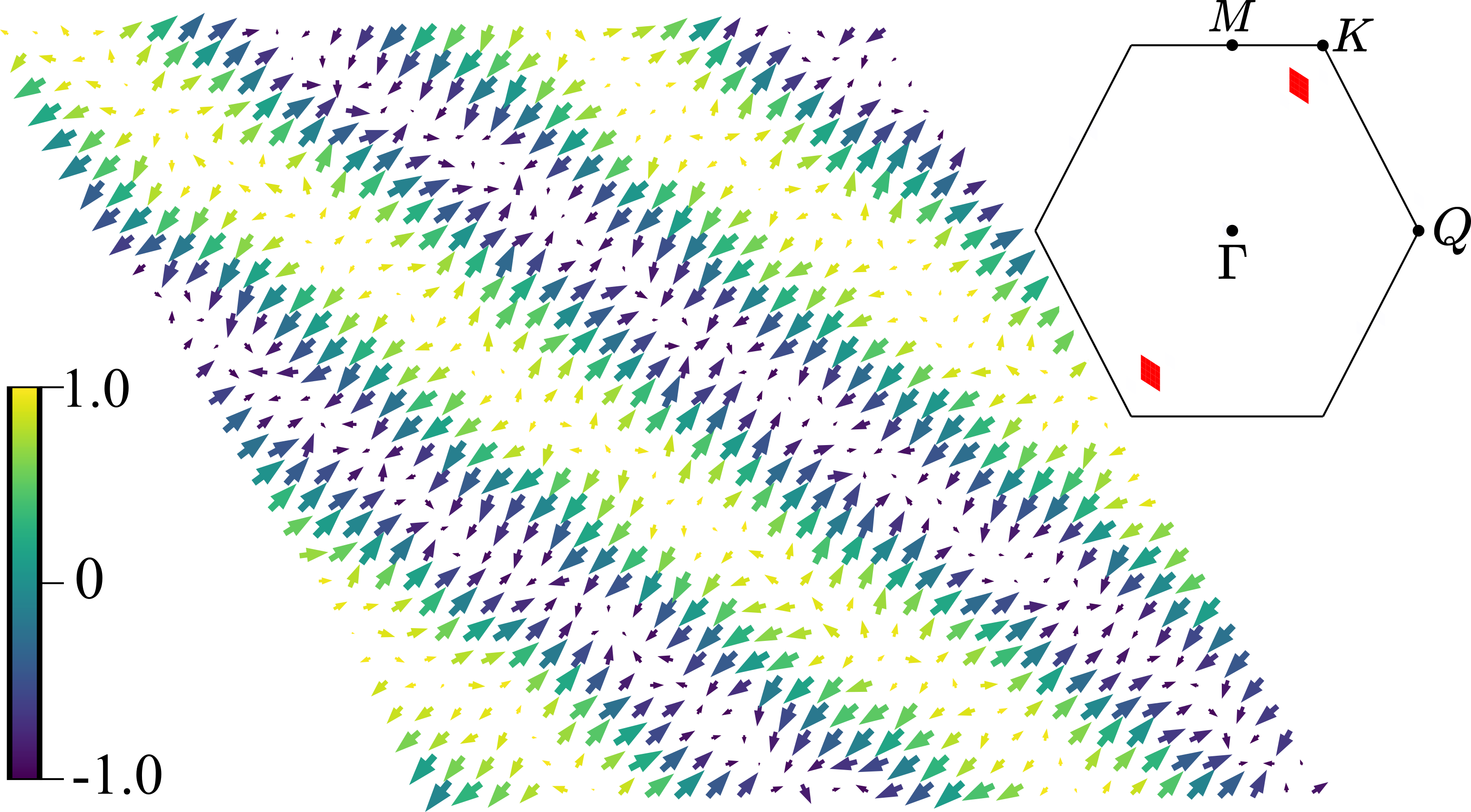}
  \caption{\textbf{Spin spiral configuration at zero field.} 
  Snapshot of the spin spiral configuration of CrInTe$_3$ on a triangular lattice at $B=0$ T. The $x$ and $y$ components of the spins are represented by arrows, while the $z$ component is color-coded. A $30 \times 30$ section of the simulated lattice is shown for clarity. The inset depicts the static structure factor of the spiral state in the first Brillouin zone (1BZ).}
  \label{fig:spinspiral_ssf}
\end{figure}

We proceed to analyze the numerical results associated with the spin spiral state of CrInTe$_3$ [see Fig. \ref{fig:spinspiral_ssf} (a)] at zero magnetic field ($B = 0$) T, we first analyze the magnetic state by computing the static structure factor (SSF):

\begin{eqnarray}
  \centering
  S_f({\bf q}) &=& S^{x}_f({\bf q}) + S^{y}_f({\bf q}) + S^{z}_f({\bf q}), \nonumber \\
  S^{\mu}_{f}({\bf q}) &=& \frac{1}{N^2} \bigg \langle \sum_{ij} S^{\mu}_i S^{\mu}_j~ e^{-{\rm i}{\bf q} \cdot ({\bf r}_i - {\bf r}_j)} \bigg \rangle
 \label{eq: SSF}
\end{eqnarray}

\noindent
where \( \mu = x, y, z \) denotes the components of the spin vector, and ${\bf r}_i$ is the position vector of spin ${\bf S}_i$. The spin spiral state is characterized by a dominant Bragg peak at wave vectors $\pm {\bf q}$ in the first Brillouin zone (1BZ) [see Fig. \ref{fig:spinspiral_ssf} inset]. Here, $\bf q$ represents the spin modulation wave vector, which describes the periodicity of the spiral pattern in reciprocal space. The specific value of $\bf q$ is determined by the competition between exchange interactions and Dzyaloshinskii–Moriya interactions (DMI), which stabilize the spiral magnetic order. In the case of a single-Q spiral state, the system spontaneously selects one of three equivalent high-symmetry points, denoted as $K$, within the 1BZ. These characteristic Bragg peaks are observable in neutron scattering experiments, providing direct evidence of the underlying spin modulation.

Now, we study the effect of this spiral state on the superconducting layer by solving the Hamiltonian in Eq. \eqref{eq:Ham_exchange_SC_Numbu}. We use the exact diagonalization method to get the eigenvalue spectrum and the local density of states (LDOS). The LDOS at lattice site index $i$ and at energy $E$ is defined as,
\begin{eqnarray} \label{eq:LDoS}
    \text{LDOS}(i,E) &=& \sum_{m, \alpha} \Big[ \langle \Psi_0|c_{i\alpha}| \Psi_m \rangle^2 \delta(E-E_m+E_0) + \langle \Psi_0|c^{\dagger}_{i\alpha}| \Psi_m \rangle^2 \delta(E-E_0+E_m) \Big]
\end{eqnarray}

\begin{figure*}[ht]
  \centering
  \includegraphics[width=0.9\textwidth,clip=true]{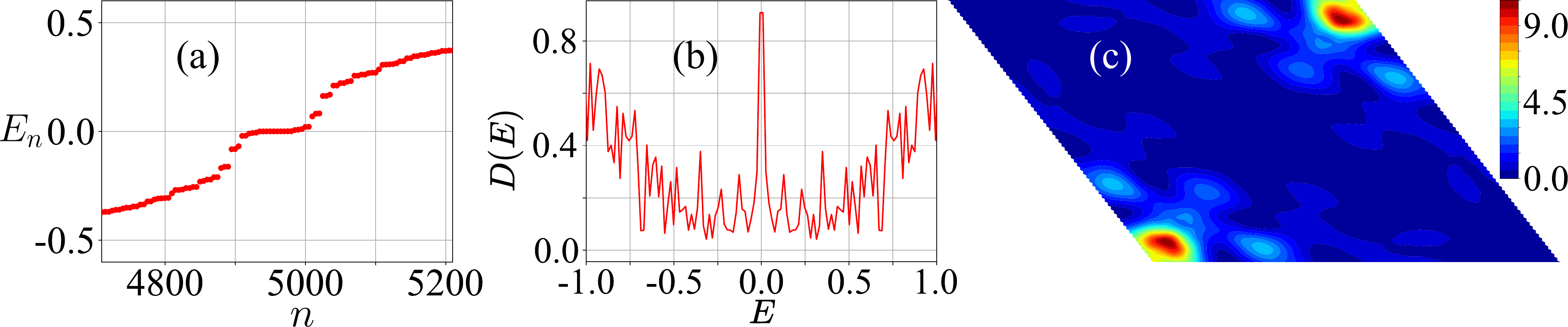}
  \caption{\textbf{Electronic properties at $J = 1.5$ and $\mu = 1.0$.} 
  (a) Eigenvalue spectrum obtained from exact diagonalization of the Hamiltonian (Eq.~\ref{eq:Ham_exchange_SC_Numbu}) coupled with a spiral spin phase. 
  (b) Total density of states (DOS), showing a zero-energy spectral peak, which numerically indicates the presence of MZM. 
  (c) Local density of states (LDOS) distribution at $E = 0$, highlighting the spatial localization of Majorana modes in the system.}
  \label{fig:eval_ldos_tdos}
\end{figure*}

\noindent
At $J = 1.5, \mu = 1.0$, we have illustrated the eigen spectrum $E_n$ as a function of the eigenstate index $n$ in the Fig. \ref{fig:eval_ldos_tdos}(a). Fig. \ref{fig:eval_ldos_tdos}(a) shows a zero-energy mode in the eigen spectrum $E_n$, which is a definite signature of the stabilization of Majorana Zero Mode (MZM). This MZM state is more evident in the LDOS plot associated with $E = 0$, shown in Fig. \ref{fig:eval_ldos_tdos}(c), obtained by utilizing open boundary conditions (OBC). The strong localization of these states at the diagonal corners of the 2D triangular lattice corroborates the topological nature of the system. In Fig. \ref{fig:eval_ldos_tdos} (b), the total density of states (DOS) exhibits a peak at zero energy, a feature reminiscent of the zero-bias peak observed in experimental studies involving differential conductance (dI/dV) spectroscopy  measurements on Majorana modes \cite{kong2021, Flensberg2017}. The DOS is computed as,

\begin{eqnarray}
  \text{DOS} (E) &=& \sum_n \delta (E - E_n) = \sum_n \frac{1}{\pi} \frac{\Gamma}{(E-E_n)^2 + \Gamma^2}
\end{eqnarray}

\noindent
where, the Lorentzian broadening parameter $\Gamma$ is taken to be 0.01.

\begin{figure}[ht]
\centering
  \includegraphics[width=0.7 \columnwidth,angle=0,clip=true]{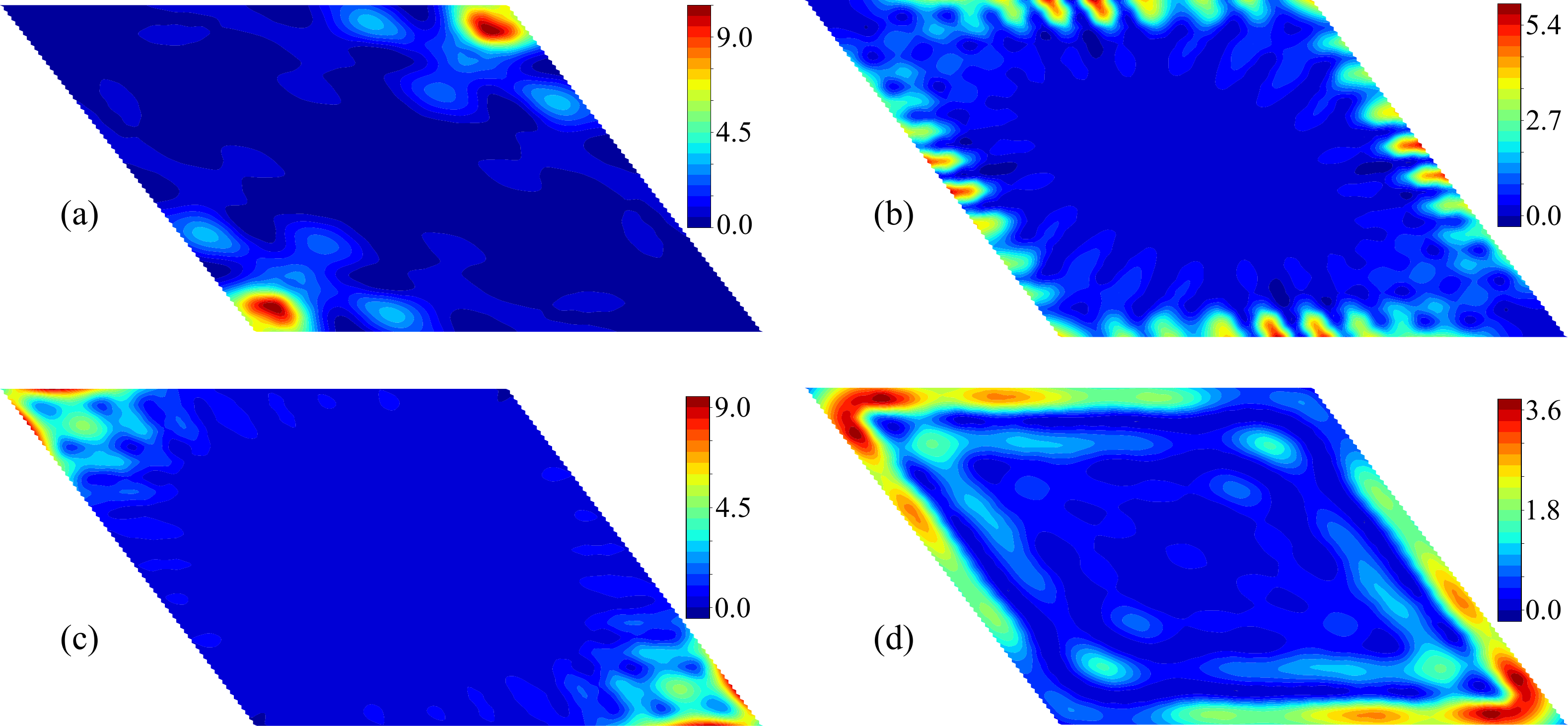}
  \caption{\textbf{Evolution of zero-energy diagonal corner state to edge state}. Zero-energy LDOS at $J = 1.5$ with (a) $\mu = 0.0$, (b) $\mu = 3.5$, (c) $\mu = 4.0$, and (d) $\mu = 4.5$}
  \label{fig:ldos_evo}
\end{figure}

Additionally, we observe the unique change in the LDOS profile within the topological region. We present the zero-energy LDOS distribution at $J = 1.5$ with varying $\mu$ [Fig. \ref{fig:ldos_evo}]. At $\mu = 0.0$, the localized Majorana modes appear in the LDOS at the diagonal corners. With increasing $\mu$, we observe the delocalization of these corner modes along the edges of the lattice. At $\mu = 4.0$, the localized zero-energy Majorana mode shifts to the other corners along opposite diagonal. However, at $\mu = 4.5$, the corresponding LDOS [see Fig. \ref{fig:ldos_evo} (b)] shows Majorana edge states across the boundary on the lattice system. A hybrid-order topological superconducting (TSC) phase, akin to this one, has been identified in the Benalcazar-Bernevig-Hughes (BBH) model. This phase emerges under conventional spin-singlet $s$-wave superconductivity, with the additional capability of adjusting an in-plane magnetic field \cite{Rodrigo2024}. This well-known extended Majorana edge mode (MEM) is protected by a bulk energy gap, which ensures robustness against local perturbations and thermal excitations. The wavefunctions of the Majorana modes decay exponentially into the bulk, ensuring that the overlap between modes is negligible, thus maintaining their topological protection and non-Abelian statistics.

\begin{figure}[ht]
  \centering
  \includegraphics[width=0.5 \columnwidth,angle=0,clip=true]{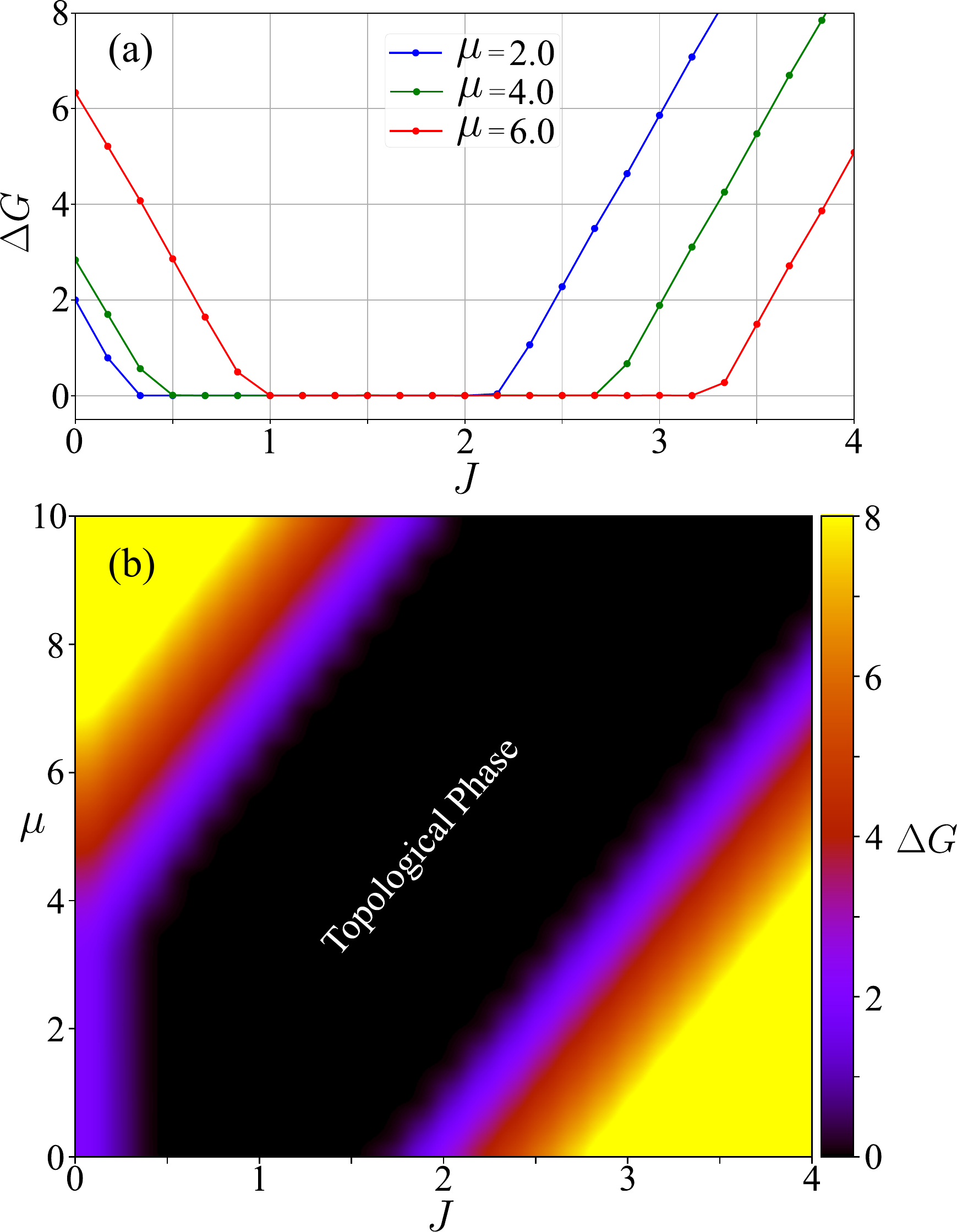}
  \caption{\textbf{$\mu$ vs $J$ phase diagram}. (a) The variation of bulk-gap $\Delta G$ with $J$ for different $\mu$ values. (b) The bulk-gap $\Delta G$ profile is shown in the $J - \mu$ plane. The dark region corresponds to the TSC phase.}
  \label{fig:bulkgap_pd}
\end{figure}

Finally, we summarize our results on TSC as a phase diagram in the $J - \mu$ plane. The emergent TSC state is quite robust in the $J - \mu$ phase plane, providing an excellent platform for experimental realization in various materials. It becomes important to mention that the value of the coupling constant, $J$, is of prime importance in the phase transition between trivial superconducting (SC) and topological superconducting (TSC) phases. However, in many cases, the value of $J$ is not easily ascertained for such materials. To identify the TSC region in the parameter space, we employ Periodic boundary condition (PBC) analysis. PBC plays a key role in identifying the regime of parameters that support the emergence of Majorana zero modes (MZMs) by focusing on the behavior of the bulk gap, $\Delta G = |E_2 - E_1|$, where $E_1$ and $E_2$ are the two lowest positive-energy eigenvalues of the bulk BdG spectrum from Eq. \ref{eq:Ham_exchange_SC_Numbu}. PBC ensures that the analysis reflects the intrinsic bulk properties of the system, free from edge effects that could otherwise obscure the topological transitions. This approach is crucial for characterizing the topological phases in a controlled and systematic manner. In Fig. \ref{fig:bulkgap_pd} (a), we show the bulk gap $\Delta G$ plotted as a function of $J$ for $\mu = 2.0, 4.0, 6.0$. In the topological regime, the bulk gap closes ($\Delta G = 0$), associated with the appearance of Majorana zero modes. Figure \ref{fig:bulkgap_pd} (b) displays $\Delta G$ in the $J - \mu$ plane. The dark strip of $\Delta G = 0$ corresponds to the gapless TSC regime, while the $\Delta G > 0$ region represents a gapped trivial superconducting phase.

\subsection*{Electronic localization effect}

\begin{figure}[ht]
\centering
  \includegraphics[width=0.5 \columnwidth,angle=0,clip=true]{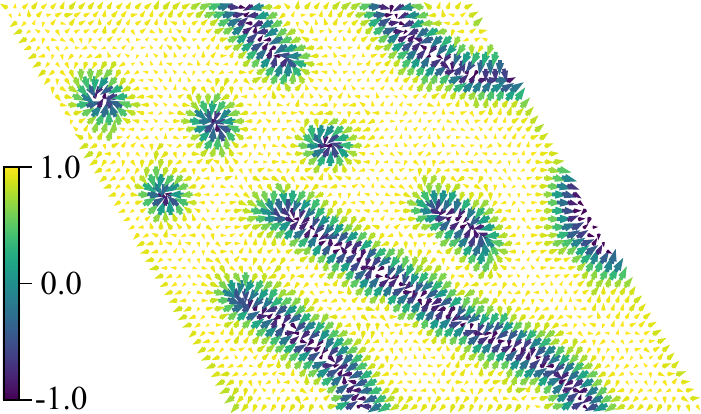}
  \caption{\textbf{Spin spiral configuration at $B=10$ T.} Snapshot of the mixed spin state configuration of CrInTe$_3$ on a $50 \times 50$ triangular lattice. The arrow indicates the $x$, $y$ components of the spins, and the colorbar denotes the $z$ component.}
  \label{fig:spinspiral_B10T}
\end{figure}

After the emergence of topological superconductivity in the presence of the magnetic spiral state at $B=0$ T -- evidenced by the induced zero-energy corner and edge states in the LDOS -- we now investigate what happens when an external magnetic field is applied. In particular, we focus on the spin textures at $B=10$ T and the corresponding changes in the electronic behavior. The spin configuration at $B = 10$ T, obtained from Monte Carlo simulations, is shown in Fig. \ref{fig:spinspiral_B10T}, where it is seen that the degeneracies of the spiral states get lifted \cite{Kathyat2020, Mukherjee2021}. The system settles into a mixed state, with the emergence of isolated skyrmions coexisting with the spiral state background. Experimentally, such exotic magnetic states are not uncommon under external fields and have been observed in various chiral and frustrated magnets \cite{grebenchuk2024, luo2023}.

\begin{figure}[ht]
  \centering
  \includegraphics[width=0.7 \columnwidth,angle=0,clip=true]{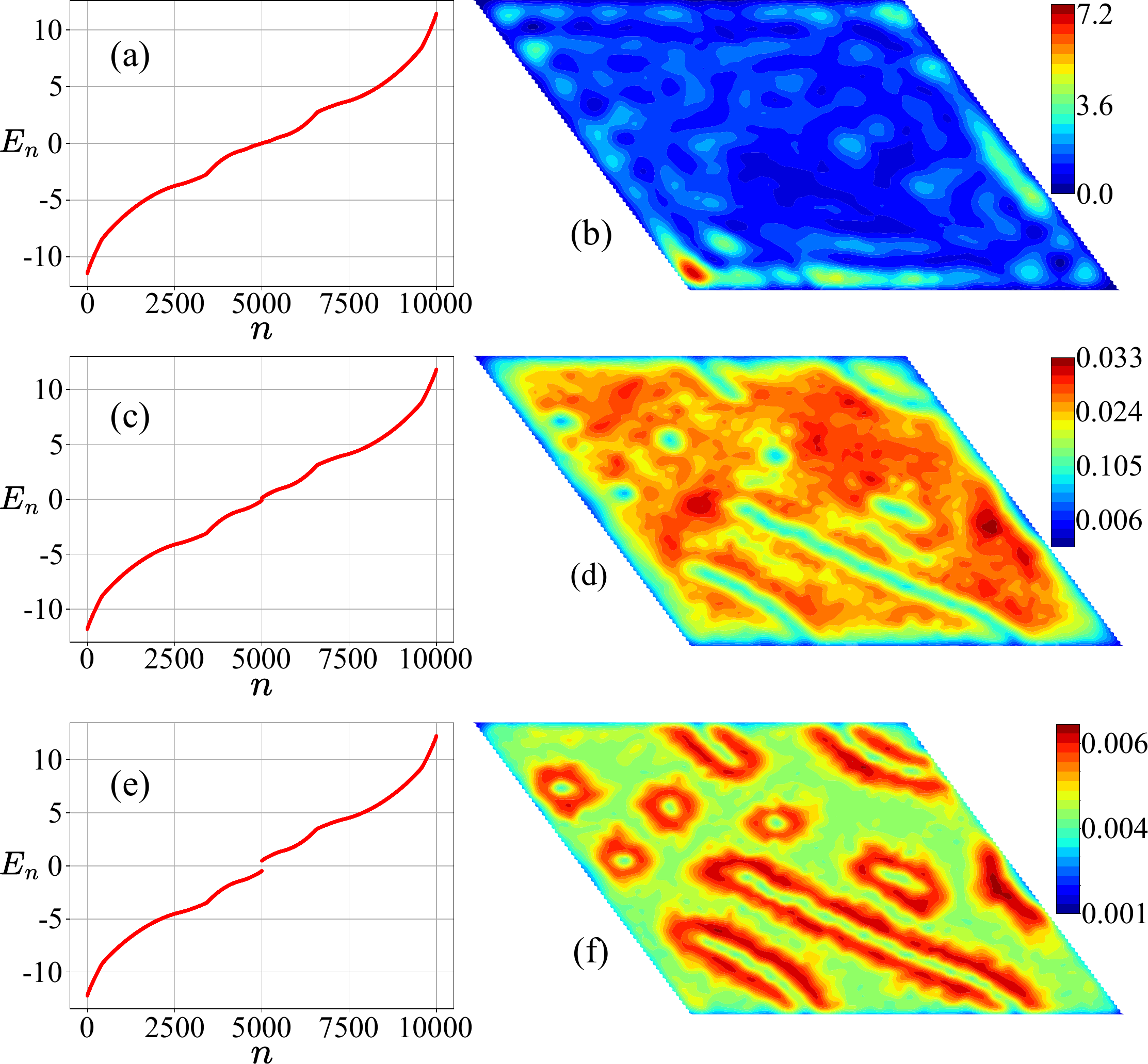}
  \caption{\textbf{Electronic localization effect}. Eigen spectrum and LDOS at fixed $J=1.5$ varying $\mu=4.0$ (a, b), $\mu=4.4$ (c, d), $\mu=4.8$ (e, f)}
  \label{fig:LDoS_J_2.0}
\end{figure}

An immediate consequence of the mixed magnetic state is the disappearance of the zero-energy Majorana modes. These modes rely on topological protection inherent in homogeneous or well-defined spiral textures; once the spin configuration becomes inhomogeneous and disordered at the mesoscopic scale, the topological superconducting phase is lost. In other words, the high degree of magnetic disorder introduced by coexisting skyrmions and spirals disrupts the delicate band topology needed for Majorana bound states. Therefore, the topological signature in the LDOS at zero energy vanishes under this external field. Instead, we observe a unique electronic localization effect that manifests in the low-energy density of states (LDOS). Focusing on $J=1.5$, Fig.~\ref{fig:LDoS_J_2.0}(a)–(f) presents the LDOS for the lowest-energy particle sector as a function of the chemical potential $\mu$.

At small $\mu=4.0$, the system behaves much like a metallic magnet, exhibiting a gapless spectrum and an almost uniform LDOS homogeneously distributed across the lattice [Fig.\ref{fig:LDoS_J_2.0}(a), (b)]. Increasing $\mu$ to 4.4 pushes the system towards partial localization, as electrons begin to cluster around magnetic domain boundaries [Fig.\ref{fig:LDoS_J_2.0}(c), (d)]. Finally, at $\mu=4.8$, the LDOS exhibits prominent spatial inhomogeneities, with strong localization concentrated around the edges of skyrmions and spiral domains [Fig.~\ref{fig:LDoS_J_2.0}(e), (f)]. A hard gap near zero energy in the eigenspectrum accompanies this localization, confirming the insulating nature of the state. Hence, as the chemical potential moves further from the optimal region for topological superconductivity, the electronic wave functions become spatially confined to regions of high magnetic inhomogeneity.

This localization arises from the complex interplay of strong magnetic disorder (caused by mixed chiral spin textures) and the electronic degrees of freedom, reminiscent of disorder-driven mechanisms in other model systems \cite{Ying2018, miranda1999localization}. Similar to how nonuniform spin states can induce localization in double-exchange or Anderson-type models, the real-space variation of the spins in our M/SC system at finite field creates effective scattering or potential wells for the electrons. As a result, transport is suppressed, and the system transits into an insulating regime despite the absence of conventional charge-disorder potential. This phenomenon highlights the central role of magnetic textures in controlling electronic band structure and underscores why the topological superconducting phase collapses under field-driven magnetic inhomogeneity.

In summary, although the spiral magnetism at $B=0$ T supports robust Majorana zero-energy modes and a topologically nontrivial superconducting phase, applying a high magnetic field drives the system into a mixed spiral–skyrmion state, destroying the topological order and giving way to strongly localized electronic states. This transition reflects a broader theme in frustrated magnetic systems, where external tuning parameters—like fields or doping—can drastically alter magnetic textures and, in turn, the electronic ground state.

\section*{Conclusion}

We have shown that two-dimensional magnet–superconductor (M/SC) heterostructures with noncoplanar magnetic textures can host robust topological superconductivity, supporting Majorana zero modes (MZMs). Concentrating on monolayer CrInTe$_3$, our combined density functional theory (DFT) and many-body analyses reveal that chiral or spiral spin configurations, when interfaced with an $s$-wave superconductor, produce nontrivial topological phases characterized by zero-energy modes at corners or edges of the lattice. The key mechanism is the interplay of geometrical frustration, Dzyaloshinskii–Moriya interactions, and conventional superconducting pairing, which together place the system in Altland-Zirnbauer Class $D$, breaking time-reversal symmetry yet preserving particle-hole symmetry.

Using a generic lattice model, we demonstrated the emergence of MZMs through numerical results for the eigenvalue spectrum and local density of states (LDOS), and further mapped out a topological superconducting (TSC) phase diagram in the $J-\mu$ plane by monitoring the bulk gap $\Delta G$. This gap closure aligns with the 2D Kitaev model signature, offering a practical route for realizing Majorana modes with emergent $p$-wave pairing in CrInTe$_3$-based hybrids. Importantly, applying an external magnetic field was found to destabilize the spiral order, driving the system into a mixed skyrmion–spiral state and thereby destroying the Majorana excitations. The quasiparticles instead become strongly localized in regions of high magnetic disorder, giving rise to an insulating state and underscoring the crucial role of magnetic inhomogeneity in tuning electronic topology.

From an experimental perspective, conventional superconductors such as Nb(110) ($\Delta_s \approx 1.51$ meV) can provide a suitable substrate for CrInTe$_3$, and the model parameters $t \approx 2.517$ meV, $J \approx 3.775$ meV, and $\mu \approx 2.517$ meV lie in a range that should permit robust TSC phases in real materials \cite{Odobesko2019, Odobesko2020}. Although the kinetic energy in actual systems can exceed the superconducting gap by two to three orders of magnitude, our results show that finite lattice geometries and magnetic tailoring can still stabilize the topological phase.

Overall, these findings present a blueprint for engineering and detecting Majorana modes in realistic 2D magnets, especially using advanced spectroscopy and scanning tunneling microscopy (STM). By clarifying the microscopic conditions for the emergence and destruction of topological superconductivity, our work provides guiding principles for designing robust quantum platforms. The capacity to control Majorana modes in tailor-made magnetic textures holds exciting potential for fault-tolerant quantum computation and topological quantum information processing.

%\bibliography{bibtex}

\section*{Acknowledgements}
B.S. acknowledges financial support from Swedish Research Council (grant no. 2022-04309), STINT Mobility Grant for Internationalization (grant no. MG2022-9386) and DST-SPARC, India (Ref. No. SPARC/2019-2020/P1879/SL).
The computations were enabled by resources provided by the National Academic Infrastructure for Supercomputing in Sweden (NAISS) at NSC and PDC (NAISS 2024/3-40) partially funded by the Swedish Research Council through grant agreement no. 2022-06725 and at UPPMAX (NAISS 2025/2-203). B.S. also acknowledges the allocation of supercomputing hours granted by the EuroHPC JU Development Access call in LUMI-C supercomputer (grant no. EHPC-DEV-2024D04-071) in Finland.
D.W. acknowledges financial support from the Science and Technology Development Fund from Macao SAR (Grant No. 0062/2023/ITP2) and the Macao Polytechnic University (Grant No. RP/FCA-03/2023).
T.N. would like to thank NFSG from BITS Pilani-NFSG/HYD/2023/H0911

\section*{Author contributions statement}
A.M., B.S, and D.W. formulated the problem. A.M., F.Z., and S.E. carried out the simulations and theoretical analysis. A.M, D.W. wrote the manuscript with inputs from F.Z., S.E., T.N., and B.S. All authors reviewed the manuscript.

\section*{Competing interests}
The authors declare no competing interests.

\section*{Figures \& Tables}

% \begin{table}[ht]
% \centering
% \begin{tabular}{|l|l|l|l|l|l|}
% \hline
%  & $J_1$ & $J_2$ & $|\textbf{D}_1|$ & $|\textbf{D}_2|$ & $K$ \\
% \hline
% CrInTe$_3$ & 4.75 & 0.179 & 1.86 & 1.20 & 2.42 \\
% \hline
% \end{tabular}
% \caption{Calculated Heisenberg exchange interaction ($J_1$, $J_2$), Dzyaloshinskii-Moriya interaction ($\textbf{D}_1$, $\textbf{D}_2$), and single ion anisotropy energy ($K$) for CrInTe$_3$, unit in meV.}
% \label{tab:m_parameters}
% \end{table}

\begin{table}[ht]
\centering
\renewcommand{\arraystretch}{1.2} % Adjust row height
\setlength{\tabcolsep}{10pt}     % Adjust column spacing
\begin{tabular}{lcccccc}
\hline
Material      & $\mathcal{J}_1$ (meV) & $\mathcal{J}_2$ (meV) & $|\textbf{D}_1|$ (meV) & $|\textbf{D}_2|$ (meV) & $K$ (meV) \\
\hline
CrInTe$_3$    & 4.75        & 0.179       & 1.86                   & 1.20                   & 2.42     \\
\hline
\end{tabular}
\caption{\textbf{Magnetic parameters for CrInTe$_3$.} The parameters include nearest-neighbor exchange interaction ($\mathcal{J}_1$), next-nearest-neighbor exchange interaction ($\mathcal{J}_2$), Dzyaloshinskii–Moriya interaction magnitudes ($|\textbf{D}_1|$ and $|\textbf{D}_2|$), and single-ion anisotropy ($K$).}
\label{tab:m_parameters}
\end{table}

 % \begin{figure}[ht]
 %  \centering
 %  \includegraphics[width=0.5 \columnwidth,angle=0,clip=true]{fig1_schematic.pdf}
 %  \caption{The schematic setup of our model consists of a two-dimensional (2D) triangular lattice of CrInTe$_3$ with a spin spiral state, placed on the surface of an $s$-wave superconductor.}
 %  \label{fig:schematic}
 % \end{figure}

 % \begin{figure}[ht]
 %  \centering
 %  \includegraphics[width=0.5 \columnwidth,angle=0,clip=true]{fig2_spinconfig_B0T.png}
 %  \caption{(a) Snapshot of the spin spiral configuration of CrInTe$_3$ on triangular lattice at $B=0$ T. The $x$ and $y$ components of the spins are indicated by the arrow while the $z$ component is color-coded. For clarity, we only display a $30 \times 30$ section of the simulated lattice. The inset represents the static structure factor of the spiral state in the 1BZ.}
 %  \label{fig:spinspiral_ssf}
 % \end{figure}

%  \begin{figure*}[ht]
%   \centering
%   \includegraphics[width=0.9 \columnwidth,angle=0,clip=true]{fig3_J_1.5_mu_1.0.pdf}
%   \caption{Electronic results at $J = 1.5, \mu = 1.0$: (a) Eigen spectrum from exact diagonalization of the Hamiltonian \ref{eq:Ham_exchange_SC_Numbu} in coupling with spiral spin phase, (b) total density of states, (c) the LDOS distribution at $E = 0$}
%   \label{fig:eval_ldos_tdos}
% \end{figure*}

\end{document}